\documentclass[pra,twocolumn,showpacs,preprintnumbers,amsmath,amssymb,superscriptaddress,unsortedaddress]{revtex4}

\usepackage{graphicx}
\usepackage{dcolumn}
\usepackage{bm}

\begin{document}

\preprint{}

\title{Adaptive quantum teleportation}

\author{Joanna Mod{\l}awska}

\affiliation{Faculty of Physics, Adam Mickiewicz University, ul. Umultowska 85, 61-614 Pozna\'{n}, Poland}

\author{Andrzej Grudka}

\affiliation{Faculty of Physics, Adam Mickiewicz University, ul. Umultowska 85, 61-614 Pozna\'{n}, Poland}

\affiliation{Institute of Theoretical Physics and Astrophysics,
University of Gda\'{n}sk, ul.  Wita Stwosza 57, 80-952 Gda\'{n}sk, Poland}

\affiliation{National Quantum Information Centre of Gda\'{n}sk, ul. W{\l}adys{\l}awa Andersa 27, 81-824 Sopot, Poland}

\date{\today}

\begin{abstract}
We consider multiple teleportation in the Knill-Laflamme-Milburn (KLM) scheme. We introduce \emph{adaptive teleportation}, i.e.,  such that the choice of entangled state used in the next teleportation depends on the results of the measurements performed during the previous teleportations. We show that adaptive teleportation  enables  an increase in the probability of faithful multiple teleportation in the KLM scheme.  In particular if a qubit is to be teleported more than once then it is better to use nonmaximally entangled states than maximally entangled ones in order to achieve the highest probability of faithful teleportation.
\end{abstract}

\maketitle

Quantum teleportation is one of the basic primitives of quantum information theory \cite{Bennett5}. Together with entanglement distillation,  it allows  sending  quantum information reliably even through quantum channels  of  vanishing zero-way quantum capacity \cite{Bennett3}. It can be also very useful in implementation of controlled quantum gates \cite{Gottesman}. The latter application is very important in linear optical quantum computation, where controlled quantum gates are implemented by means of quantum teleportation \cite{Knill}. If Alice wants to teleport a qubit to Bob, then she performs complete Bell measurement on a qubit to be teleported and a qubit from the maximally entangled pair which she shares with Bob. After learning the result of Alice's measurement Bob performs one of four unitary operations. However, sometimes the parties cannot perform complete Bell measurement. Such  a  situation naturally occurs in quantum information processing with linear optics \cite{Lutkenhaus}. Partially due to this fact,  the first experimental demonstration of quantum teleportation succeeded only in 25\% \cite{Bouwmeester}. Knill, Laflamme and Milburn presented a protocol which enables linear optical teleportation with high probability \cite{Knill}. In their protocol a qubit is encoded in superposition of vacuum and one photon state. The protocol also uses a special $N$-photon entangled state. The probability that quantum teleportation succeeds is equal to $1-\frac{1}{N+1}$ and when it happens,  the  fidelity of the teleported qubit with the original qubit is equal to $1$. Recently the protocol was generalized to polarization encoding  \cite{Spedalieri}.  In \cite{Grudka} we  have  found that the maximally entangled state gives the highest probability of successful single teleportation in the KLM protocol. 

The situation is quite different when one considers a chain of linear optical teleportations, i.e., Alice teleports a qubit to Bob, then Bob teleports the qubit, which he received from Alice, to Charlie and so on. If we assume that each party performs KLM protocol and uses \emph{identical} entangled state,  then for  a  sufficiently large number of teleportations the  nonmaximally entangled states give the highest probability of successful teleportation \cite{Modlawska}. However, in order to achieve faithful teleportation one has to correct errors after the last teleportations by performing local filtering.

In this paper we show, how one can further increase  the  probability of faithful multiple linear optical teleportation. We introduce adaptive teleportation. In this protocol the choice of entangled state used in the next teleportation depends on the results of measurements obtained during the previous teleportations. We show that the simple version of this protocol enables  an  increase  in the  probability of two teleportations. Moreover, the protocol has  a built-in error correction, i.e., the last teleportation corrects errors. We also compare  the  adaptive teleportation with  the  multiple teleportation when one uses identical entangled states in all teleportations.

In order to illustrate  the  adaptive teleportation we consider three party scenario in which Alice teleports a qubit to Bob and then Bob teleports the  qubit received  to Charlie (Fig. 1). We assume that each party performs  the  KLM protocol and uses  the  entangled state with $N$ vertically polarized photons and $N$ horizontally polarized photons.  The  entangled state used in each teleportation can be different. Moreover, Bob and Charlie can choose  the  entangled state which they use in teleportation after learning  the result of Alice's measurement. Our aim will be maximization of probability of faithful multiple teleportation. The teleportation is faithful if fidelity of the teleported qubit after the last teleportation with the original qubit is equal to 1.

We assume that Alice and Bob share  the  entangled state of the form
\begin{equation}
|t_{1}\rangle=\sum_{i=0}^{N}c_{1}(i)|V\rangle^{i}|H\rangle^{N-i}|H\rangle^{i}|V\rangle^{N-i},
\label{eq:1}
\end{equation}
where $|V\rangle^{i}$ stands for $|V\rangle_{1}|V\rangle_{2}...|V\rangle_{i}$, i.e., one vertically polarized photon in each of the subsequent modes. Similarly,  $|H\rangle^{N-i}$ stands for $|H\rangle_{i+1}|H\rangle_{i+2}...|H\rangle_{N}$ i.e., one horizontally polarized photon in each of the subsequent modes.  The  first $N$ modes are held by Alice and  the  next $N$ modes are held by Bob. If all amplitudes $c_1(i)$ are the same,  then the state is maximally entangled. In order to teleport a qubit in  the  state $|\psi\rangle=\alpha|H \rangle+\beta|V\rangle$ Alice applies to the input mode and  the  first $N$ modes of  the  entangled state  the  $N+1$-point quantum Fourier transform
\begin{eqnarray}
& F_{N} (h_{k}^{\dagger})=\frac{1}{\sqrt{N+1}} \sum_{l_{k}=0}^{N} \omega^{k l_{k}} h_{l_{k}}^{\dagger},
\nonumber \\ 
& F_{N} (v_{k}^{\dagger})=\frac{1}{\sqrt{N+1}} \sum_{l_{k}=0}^{N} \omega^{k l_{k}} v_{l_{k}}^{\dagger},
\label{eq:2}
\end{eqnarray}
where $h_{k}^{\dagger}$ ($v_{k}^{\dagger}$) is  the  creation operator for a horizontally (vertically) polarized photon in mode $k$ and $\omega = e^{\frac{i 2 \pi}{N+1}}$. Next, Alice measures  the  total number of horizontally and vertically polarized photons in each of these modes. If the sum of  vertically polarized photons detected is $m$ ($0<m<N+1$), then the modified state of the qubit is found in the $N+m$-th mode. After phase correction the state of the qubit is
\begin{equation}
|\psi\rangle=\frac{1}{\sqrt{p(m)}}(\alpha c_{1}(m)|H \rangle+\beta c_{1}(m-1)|V\rangle),
\label{eq:3}
\end{equation}
where 
\begin{equation}
p(m)=|\alpha c_{1}(m)|^2+|\beta c_{1}(m-1)|^2
\label{eq:4}
\end{equation}
is probability that the sum of vertically polarized photons detected is equal to $m$.
Next, Bob teleports a qubit which he received from Alice to Charlie. We assume that Bob and Charlie share several entangled states of the form 
\begin{equation}
|t_{2; m}\rangle=\sum_{i=0}^{N}c_{2; m}(i)|V\rangle^{i}|H \rangle^{N-i}|H \rangle^{i}|V\rangle^{N-i}.
\label{eq:5}
\end{equation} 
The amplitudes $c_{2; m}(i)$ can be different from  $c_{1}(i)$. Moreover, they can also be different for each entangled state shared by Bob and Charlie. The choice of  the  entangled state used in the teleportation depends on the result of Alice's measurement. After  the  choice of  the  entangled state,  Bob applies,   the  $N+1$-point quantum Fourier transform  to the input mode and  the  first $N$ modes of \emph{chosen} entangled state  and measures the total number of horizontally and vertically polarized photons in each of these modes.   If the sum of vertically polarized photons detected is $n$ ($0<n<N+1$) then the modified state of the qubit is found in the $N+n$-th mode. After  the  phase correction the state of the qubit is
\begin{eqnarray}
& |\psi\rangle= \frac{1}{\sqrt{p(m,n)}}\nonumber\\
& (\alpha c_{1}(m) c_{2; m}(n)|H \rangle+\beta c_{1}(m-1) c_{2; m}(n-1)|V \rangle),
\label{eq:6}
\end{eqnarray}
where
\begin{eqnarray}
& p(m,n)=p(n|m)p(m)= \nonumber\\
& =|\alpha c_{1}(m) c_{2; m}(n)|^2+|\beta c_{1}(m-1) c_{2; m}(n-1)|^2
\label{eq:7}
\end{eqnarray}
is  the  probability  that the sum of vertically polarized photons detected during the first teleportation is equal to $m$ and the sum of vertically polarized photons detected during the second teleportation is equal to $n$.
\begin{figure}
\includegraphics [width=6truecm]{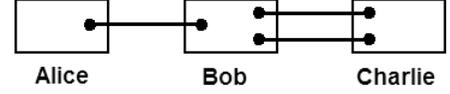}
\caption{\label{fig:1} Adaptive quantum teleportation. Bob and Charlie share two different entangled states. For some results of Alice's measurement Bob teleports the qubit to Charlie using the "upper" entangled state and for some results of Alice's measurement Bob teleports the qubit to Charlie using the "lower" entangled state.}
\end{figure}
We require that the qubit is teleported faithfully, i.e., its state after the second teleportation is equal to
\begin{equation}
|\psi\rangle=\alpha |H \rangle+\beta |V \rangle,
\label{eq:8}
\end{equation}
and hence the amplitudes $c_{1}(m)$ and $c_{2; m}(n)$ should satisfy the condition
\begin{equation}
c_{1}(m)c_{2; m}(n)=c_{1}(m-1)c_{2; m}(n-1). 
\label{eq:9}
\end{equation}
If  $c_{1}(m)<c_{1}(m-1)$ ($c_{1}(m)>c_{1}(m-1)$),  then the amplitudes $c_{2; m}(n)$ should form an increasing (decreasing) geometric sequence.

Now we make the following assumptions: 

(1) Entangled states used in both teleportations have \emph{even} $N$ \footnote{Similar calculations can be made for \emph{odd} $N$. However, if we assume that entangled state used in the first teleportation has amplitudes symmetric around $i=N/2$ then Bob and Charlie should choose one of three entangled states rather than one of two entangled states.}. 

(2)  The  entangled state used in the first teleportation has  the  amplitudes given by  the  formula
\begin{equation}
c_{1}(m)=\sqrt{a_1} (\sqrt{q_1})^{N/2-|m-N/2|},
\label{eq:11}
\end{equation}
where $q_1>1$, i.e., for  $m \leqslant N/2$ the amplitudes $c_{1}(m)$ form an increasing geometric sequence and for $m \geqslant N/2$ they form a decreasing geometric sequence.
From  the  normalization condition $\sum_{m=0}^{N}|c_{1}(m)|^2=1$ we obtain
\begin{equation}
a_1=\frac{1-q_1}{2-q_1^{N/2}(1+q_1)}.
\label{eq:12}
\end{equation}
From Eq.~(\ref{eq:9}) we obtain that  the  entangled state used in the second teleportation has  the  amplitudes given by  the  formula
\begin{equation}
c_{2; m}(n)=\sqrt{a_2}(\sqrt{q_2})^n, 
\label{eq:13}
\end{equation}
where $q_2=1/q_1$ if the number of vertically polarized photons $m$ detected during  the first teleportation is  lower or equal to $N/2$ and $q_2=q_1$ if the number of vertically polarized photons $m$ detected during the first teleportation is greater than $N/2$, i.e., Bob and Charlie choose one from  two entangled states which they share depending on the result of Alice's measurement. From  the  normalization condition $\sum_{n=0}^{N}|c_{2; m}(n)|^2=1$ we obtain
\begin{equation}
a_2=\frac{1-q_2}{1-q_2^{N+1}}.
\label{eq:14}
\end{equation} 
In order to obtain the probability of  faithful double teleportation we have to calculate the probability that both Alice and Bob register from $1$ to $N$ vertically polarized photons. If one of them registers $0$ or $N+1$ vertically polarized photons, then teleportation fails. Hence, our probability is given by the following expression
\begin{eqnarray}
& p=\sum_{m=1}^N\sum_{n=1}^Np(m,n)=\nonumber\\
& =\sum_{m=1}^N\sum_{n=1}^N|c_{1}(m) c_{2; m}(n)|^2
\label{eq:10}
\end{eqnarray}
Substituting formulae~(\ref{eq:11}) and~(\ref{eq:13}) together with~(\ref{eq:12}) and~(\ref{eq:14}) into~(\ref{eq:10})  we obtain the following expression for  the  probability of faithful double teleportation
\begin{eqnarray}
p=2 \frac{q -q^{N/2+1}}{2-q^{N/2}(1+q)} \frac{q^{N}-1}{q^{N+1}-1}.
\label{eq:15}
\end{eqnarray}
This probability should be optimized over $q$. In Table~\ref{tab:table1} we present  the  optimized probabilities for different $N$ when one performs adaptive double teleportation and when one performs double teleportation with the use of maximally entangled states. We see, that if the qubit is to be teleported at least two times, then it is better to perform adaptive teleportation than  that with the use of maximally entangled states.

It is interesting that in this simple scenario another strategy is possible which gives the same probability of faithful double teleportation. Namely, Bob always teleports the qubit to Charlie with entangled state whose amplitudes form a decreasing geometric sequence. However, if Alice detects more than $N/2$ photons during the first teleportation Bob applies NOT gate to the teleported qubit before the second teleportation and Charlie applies NOT gate to the teleported qubit after the second teleportation. Hence, we replaced the choice of entangled state by the choice of gate. 

\begin{table}
\caption{\label{tab:table1}
Probability of faithful double teleportation with maximally entangled states (second column) for different $N$. Optimal parameter $q$ (third column) and maximal probability of faithful adaptive double teleportation with entangled states whose amplitudes are given by Eqs.~(\ref{eq:11}) and~(\ref{eq:13})(fourth column).}
\begin{center}
\begin{tabular}{|c|c|c|c|} \hline
$N$ & $(\frac{N}{N+1})^2$  & $q_{opt}$ & $p_{opt}$ \\ \hline
2  &   0,444444   &  1,29663  & 0,454167  \\ \hline
4  &   0,640000   &  1,20892  & 0,652198  \\ \hline
6  &   0,734694   &  1,15822  & 0,746252  \\ \hline
8  &   0,790123   &  1,12682  & 0,800565  \\ \hline
10 &   0,826446   &  1,10569  & 0,835820   \\ \hline
\end{tabular}
\end{center}
\end{table}

Let us now compare  the  adaptive teleportation with the multiple teleportation developed in \cite{Modlawska} when one uses identical entangled states in all teleportations and error correction is made after the last teleportation. We consider only the simplest nontrivial example when one uses  the  entangled states with $N=2$.  We assume that entangled states have amplitudes symmetric around $i=1$ \footnote{It is very unlikely that entangled states which do not have amplitudes symmetric around $i=1$ are optimal. However, we were not able to prove it rigorously.}. The probability of faithful multiple teleportation is given by formula \cite{Modlawska}
\begin{eqnarray}
& p=\sum_{n_1=1}^{2}  ... \sum_{n_M=1}^{2} \nonumber \\
& \text{min}\{|c(n_1)...c(n_M)|^2, |c(n_1-1)...c(n_M-1)|^2\},
\label{eq:16}
\end{eqnarray}
where $M$ is the number of teleportations.
Substituting $q=\left|\frac{c(1)}{c(0)}\right|^2$ and using the normalization condition $\sum_{m=0}^{2}|c(m)|^2=1$ we obtain
\begin{eqnarray}
& p=\left(\frac{1}{2+q}\right)^M\sum_{i=0}^{M}\frac{M!}{i!(M-i)!} \text{min}\{q^{M-i}, q^i\}.
\label{eq:17}
\end{eqnarray}
This probability should be optimized over $q$. 

For less than six teleportations the highest probability of faithful multiple teleportation with identical entangled states is achieved for {\it maximally} entangled states. Hence, if we replace any two teleportations in a chain of two or more teleportations with  maximally entangled states by double adaptive teleportation then we obtain greater probability of faithful multiple teleportation. It should be noted that there is no contradiction because in adaptive teleportation one does not use identical entangled states.

On the other hand for six or more teleportations the highest probability of faithful multiple teleportation with identical entangled states is achieved for {\it nonmaximally} entangled states. In what follows, we show that if we replace the original nonmaximally entangled state used in the last teleportation by the adapted entangled state,  then we increase  the  probability of faithful teleportation.
Let us assume that the first sender detected $n_1$ vertically polarized photons, the second sender detected $n_2$ vertically polarized photons ... and the $M-1$-th sender detected $n_{M-1}$ vertically polarized photons. Then the state of the qubit after $M-1$ teleportations is
\begin{eqnarray}
& |\psi \rangle =\frac{1}{\sqrt{p_{M-1}(n_1, ..., n_{M-1})}} \times\nonumber\\
& \times (\alpha a(1) |H \rangle + \beta a(0) |V \rangle),
\label{eq:20}
\end{eqnarray}
where $p_{M-1}(n_1, ..., n_{M-1})=|\alpha a(1)|^2+|\beta a(0)|^2$ is probability of this event. The coefficients $a(1)$ and $a(0)$ are given by formulae
\begin{eqnarray}
& a(1)=c(n_1)...c(n_{M-1}),\nonumber \\
& a(0)=c(n_1-1)...c(n_{M-1}-1).
\end{eqnarray}
Because for six or more teleportations with identical entangled states,  the  nonmaximally entangled states give higher probability of faithful multiple teleportation, then for some results of measurements we have $a(0) \neq a(1)$. Without loss of generality we assume that $a(1)$ is \emph{greater} than $a(0)$ and write $|a(1)|^2=q|a(0)|^2$ $(q>1)$. Let $b(n_M)$ be coefficients of entangled state used in the $M$-th teleportation. If the $M$-th sender detected $n_M$ vertically polarized photons, then the state of the qubit after the $M$-th teleportation is
\begin{eqnarray}
& |\psi \rangle =\frac{1}{\sqrt{p_{M}(n_1, ..., n_{M})}} \times \nonumber \\
& \times (\alpha b(n_M) a(1) |H \rangle + \beta b(n_M-1) a(0) |V \rangle),
\end{eqnarray}
where $p_{M}(n_1, ..., n_{M})=|b(n_M)\alpha a(1)|^2+|b(n_M-1)\beta a(0)|^2$ is probability that the first sender detected $n_1$ vertically polarized photons, the second sender detected $n_2$ vertically polarized photons ... and the $M$-th sender detected $n_{M}$ vertically polarized photons. 
If $b(n_M) a(1) = b(n_M-1) a(0)$, then the qubit is in its original state. If $b(n_M) a(1) > b(n_M-1) a(0)$ then the optimal strategy to return the qubit to its original state is to perform the generalized measurement given by Kraus operators
\begin{eqnarray}
& E_S=\frac{b(n_M-1) a(0)}{b(n_M) a(1)}|H\rangle \langle H|+|V\rangle \langle V|,\nonumber \\
& E_F=\sqrt{1-|\frac{b(n_M-1) a(0)}{b(n_M) a(1)}|^2}|H\rangle \langle H|.
\end{eqnarray}
If $E_S$ is obtained as a result of the measurement then the aim is achieved. Otherwise the state of the qubit is irreversibly destroyed. Similar measurements exists if $b(n_M) a(1) < b(n_M-1) a(0)$. Probability that the first sender detected $n_1$ vertically polarized photons, the second sender detected $n_2$ vertically polarized photons ... the $M$-th sender detected $n_{M}$ vertically polarized photons and the error correction succeeded is
\begin{eqnarray}
& p_{M}(n_1, ... , n_M,S)=\nonumber\\
& =\text{min}\{|b(n_M) a(1)|^2, |b(n_{M}-1) a(0)|^2\}.
\end{eqnarray} 
In order to obtain the total probability of faithful multiple teleportation we have to sum these probabilities over the number of photons detected in each teleportation. Let us first perform summation over the number of vertically polarized photons detected in the last teleportation, i.e.,
\begin{eqnarray}
& p_{M}(n_1, ..., n_{M-1},S)= \nonumber \\
& =\sum_{n_M=1}^2\text{min}\{|b(n_M) a(1)|^2, |b(n_{M}-1) a(0)|^2\}.
\label{eq:21}
\end{eqnarray}
If we use in the $M$-th teleportation a state which has amplitudes symmetric around $n_M=1$, i.e., $b(0)=b(2)$ then  the  maximal probability $p_{M}(n_1, ..., n_{M-1},S)$ is given by  the  formula
\begin{eqnarray}
p^{\text{sym}}_{M}(n_1, ..., n_{M-1},S)=|a(0)|^2\frac{1+q}{2+q}. 
\label{eq:22}
\end{eqnarray}
On the other hand, if we use in the $M$-th teleportation a state whose amplitudes are related by $|b(n_{M}-1)|^2=q|b(n_M)|^2$, i.e., they form a \emph{decreasing} geometric sequence, then  the  probability $p_{M}(n_1, ..., n_{M-1},S)$ is given by  the  formula
\begin{eqnarray}
p^{\text{a}}_{M}(n_1, ..., n_{M-1},S)=|a(0)|^2\frac{1+q}{1+q^{-1}+q}. 
\label{eq:23}
\end{eqnarray} 
Because $q>1$ we have $p^{\text{a}}_{M}(n_1, ..., n_{M-1},S)>p^{\text{sym}}_{M}(n_1, ..., n_{M-1},S)$. 

If  $a(1)$ were \emph{smaller} than $a(0)$, then we would use in the $M$-th teleportation a state whose amplitudes  form  an \emph{increasing} geometric sequence.  The relation  between $a(1)$ and $a(0)$  depends on the results of measurements obtained during first $M-1$ teleportations and hence, the choice of entangled state used in the $M$-th teleportation also depends on the results of these measurements, i.e., we perform adaptive teleportation. Summing $p_{M}(n_1, ..., n_{M-1},S)$ over all possible results of measurements obtained during first $M-1$ teleportations,  we obtain  the  probability that the state is teleported faithfully. Because $p^{\text{a}}_{M}(n_1, ..., n_{M-1},S)$ for teleportation with an adapted entangled state is greater than the corresponding $p^{\text{sym}}_{M}(n_1, ..., n_{M-1},S)$ for  teleportation with a symmetric state (with possible exception of probabilities corresponding to the case with $a(1)=a(0)$, where they can be equal) we obtain that it is always better to perform adaptive teleportation as the last teleportation.

In conclusion,  we  have  introduced  the  adaptive teleportation,  in which  the choice of entangled state used in the next teleportation depends on the results of measurements obtained  during  the previous teleportations. We  have  also compared  the  adaptive teleportation with  the  multiple teleportation  when one uses identical entangled states in all teleportations.

\begin{acknowledgments}
One of the authors (A.G.) was partially supported by Ministry of Science and Higher
Education under Grant No. N N206 2701 33 and by the European Commission through the Integrated Project FET/QIPC SCALA.
\end{acknowledgments}

\end{document}